# A new understanding on the history of developing MRI for cancer detection

*Donald C. Chang*
*Hong Kong University of Science and Technology, Clear Water Bay, Hong Kong*
Email: *bochang@ust.hk*

**Abstract:** Science is about facts and truth. Yet sometimes the truth and facts are not obvious. For example, in the field of MRI (Magnetic Resonance Imaging), there has been a long-lasting debate about who were the major contributors in its development. Particularly, there was a strong dispute between the followers of two scientists, R. Damadian and P. Lauterbur. In this review, we carefully trace the major developments in applying NMR for cancer detection starting almost 50 years ago. The research records show that the truth was beyond the claims of either research camps. The development of NMR for cancer detection involved multiple research groups, who made critical contributions at different junctures.

## Introduction

As one of the most prevalent diseases in the world, cancer continues to be a major challenge for medical professionals. The development of NMR (nuclear magnetic resonance) for cancer detection is a major contribution of physics to the betterment of human health. MRI (magnetic resonance imaging) is one of the most powerful techniques in the diagnosis of cancer today [1–4]. However, there is a common misunderstanding in the public about MRI. It is often thought that MRI is simply a magnetic imaging technique, which is somewhat similar to CT scan, except that MRI uses the magnetic field instead of X-ray to measure the distribution of matter in the human body. This thinking is not correct. The utility of MRI is not so much for its ability to visualize the anatomical structure; its major advantage is its ability to image the physical state of cellular water, which reflects the physiological-pathological state of the biological tissue.

There was a fascinating story behind the development of NMR for cancer detection. In fact, although the MRI work started almost 50 years ago, there is still active debate today on who were the major contributors for the discovery of using NMR for cancer detection [5–10]. It has become one of the great controversies in science in modern time [5–11]. Particularly, there was a strong dispute between two scientists, P. Lauterbur and R. Damadian, on who should deserve the Nobel Prize for developing MRI [5, 7, 9–14].

Since MRI is a very important medical physics technology, we would like to clarify this confusing situation by carefully examining the research record. The key steps in developing MRI for cancer detection actually involved three parts:

- Using spin-echo NMR to measure relaxation times in cellular water.
- The discovery that the NMR relaxation times (T1, T2) and the spin diffusion coefficient (D) of cellular water are sensitive to the physiological-pathological state of the biological tissues.

- Using a magnetic field gradient for 2-D NMR scanning

It is not easy to sort out who were the major contributors for the different steps. This paper is a concise summary of the history in the early development of MRI for cancer detection. From this detailed review, one can have a more clear idea on what really happened.

## The early motivation of using NMR to study biological tissues

The application of NMR for cancer detection is a kind of serendipity in scientific research. The original purpose of using NMR to study water in biological cells was to address a debate about the physical properties of water in biological cells [15]. In the early days of physical biochemistry studies, there were two schools of thought about the physical state of cellular water. The traditional view was that the cell is a membrane bounded containment of biochemical solution [16]. So, the cellular water is no different than water contained in a test tube. Some more physically oriented scientists, however, disagreed. They thought that the cell is a highly structured ion-water-macromolecular complex. The water there is not a free fluid, but in a much more ordered physical state. [17, 18]

To settle this argument, several groups proposed to use NMR to study water in biological tissues [19–23]. There were two major types of NMR techniques: (a) the CW (continuous wave) NMR, which was mainly used by chemists for identifying chemical contents; (b) the pulsed NMR (spin-echo NMR) for measuring the spin-lattice relaxation time T1 and spin-spin relaxation time T2, which reflect the correlation times of the nuclear spins. These measurements were often used for studying the physical state of molecules. During the 1960s, some preliminary studies using both CW and pulsed NMR had been conducted on biological tissues [19–23]. There was evidence that cellular water appeared to have much reduced mobility in comparison to water in free solution.

During that time, I was a PhD student in the Department of Physics at Rice University. My advisor, Prof. H. E. Rorschach, was a collaborator of Felix Bloch, who invented the NMR technique. We built a spin-echo NMR spectrometer for studying the transport properties of quantum fluid (liquid $^3$He/$^4$He at low temperature) [24]. At later years of my PhD study, I became highly interested in biophysics. Rice University is located right next to the Texas Medical Center, where I met a young physiologist working at the Baylor College of Medicine, C. F. Hazlewood. Previously, Hazlewood had used CW NMR to study biological water [22, 23]. So, when I finished my PhD degree in 1970, I decided to collaborate with Hazlewood and converted my home-built spin-echo NMR spectrometer at Rice University to study the physical state of water protons in biological tissues.

## Discovery that the NMR relaxation times of cellular water are sensitive to the physio-pathological state of the biological tissue

Using the spin-echo NMR method, we conducted a series of studies on cellular water in different biological systems [15, 25–30]. We found that the NMR relaxation times (T1, T2) and the spin diffusion coefficient (D) of water protons are much shorter in the biological tissues in comparison to free water, suggesting that the water in the biological tissues are more ordered than the bulk water [15, 25–30]. More interestingly, the ordering of cellular water was found to be highly sensitive to the physiological state of the biological tissues. The T1, T2 and D of water protons were observed to undergo a significant change during the normal development of biological tissue. For example, when we studied the muscle cells in rats between newborn and 40 days old, we discovered that the NMR relaxation times of cellular water in the less differentiated muscle tissues were much longer than those in the more differentiated mature muscle [25, 27].(See **Table 1**).

### Table 1
Pulsed NMR measurements of relaxation times, T1 and T2, for pure water and rat skeletal muscle water (from Ref. [25])

| Sample | $T_1$ (sec) | $T_2$ (sec) |
|---|---|---|
| Pure H$_2$O | (3) 2.97 ± 0.14 | (6) 1.6 ± 0.11 |
| Mature Muscle | (6) 0.723 ± 0.049 | (6) 0.047 ± 0.004 |
| Immature Muscle | (5) 1.206 ± 0.055 | (7) 0.127 ± 0.009 |

NOTE: All values are given in seconds and are mean values ± standard error of the mean. Numbers in parentheses indicate number of samples analyzed.

This discovery had a strong implication. Since cancer development is known to be related to de-differentiation of the biological cells, our findings suggested that the relaxation times in tumor could be different from normal tissue. We collaborated with a pathologist (Dan Medina) to test this hypothesis in a mouse model of mammary tumor. This model had several advantages. First, it has three different well-defined morphological states (normal tissue, pre-neoplastic nodule, and cancer). Second, since the breast cancer is the most common cancer in woman, our study will have very strong medical relevance. Finally, if NMR can detect pre-cancer formation in the mammary tissue, it will be greatly helpful for treating the patient early to prevent the development of breast cancer.

The result of our study was very exciting; it fully confirmed our prediction [27, 28]. (See **Table 2**). In September 1971, I submitted a meeting abstract to the American Physical Society (APS) to report our experimental findings.

### Table 2
Pulsed NMR measurements of relaxation times T1 and T2 and spin diffusion coefficients D, for pure water and mouse mammary gland (from Ref. [27, 28])

| Tissue and number of samples | $T_1$ (sec) | $T_2$ (sec) | $D$ (cm²/sec × 10⁻⁵) |
|---|---|---|---|
| Pure water | 3.1 | 1.43 ± 0.27 | 2.38 ± 0.016 |
| Tumor (5) | 0.920 ± 0.047 | 0.091 ± 0.008 | 0.78 ± 0.05 |
| Nodule (5) | 0.451 ± 0.021 | 0.053 ± 0.001 | 0.44 ± 0.03 |
| Normal pregnant mammary gland (5) | 0.380 ± 0.041 | 0.039 ± 0.002 | 0.34 ± 0.04 |

During the 1972 APS March Meeting, we gave an oral report on the results of our study [27]. Our paper caught the attention of many physicists attending that meeting. As a result, our presentation became the highlight of that year's APS March Meeting. (See **Figure 1**). Subsequently, our full paper was published in June 1972 in PNAS [28].

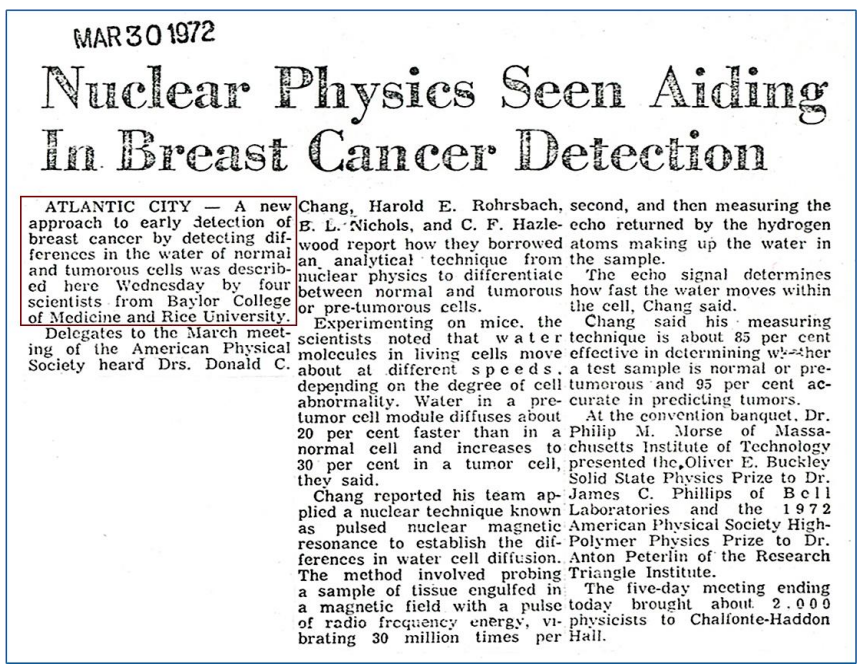

**Figure 1.** Press release of the American Physical Society (APS) for their 1972 APS March Meeting. The report of the NMR study by Chang et al was highlighted in this meeting.

In the meantime, Raymond Damadian, a medical doctor at the Downstate Medical Center in New York, was interested in the structural water hypothesis and was in contact with my collaborator (C.F. Hazelwood). He became aware of the NMR work of our team. Although Damadian had no prior training or research experience in NMR, he was very keen on seeing the important implication of our NMR study on developing tissues. He used an NMR spectrometer of the **NMR Specialties Corporation** (at West Kensington, PA) to conduct a quick measurement on 3 types of tumors. He found that the average values of T1 in some tumors were longer than those of various normal tissues (see **Table 3**). He quickly published these results in 1971 as a short report in *Science* [31].

**Table 3**
Pulsed NMR measurements of relaxation times, T1 and T2, for three different rat tumors (from Ref. [31])

| Rat No. | Weight (g) | $T_1$ | $T_2$ |
|---|---|---|---|
| *Walker sarcoma* | | | |
| 6 | 156 | 0.700 | 0.100 |
| 7 | 150 | .750 | .100 |
| 8 | 495 | .794 (0.794)* | .100 |
| 9 | 233 | .688 | |
| 10 | 255 | .750 | |
| Mean and S.E. | | 0.736 ± 0.022 | .100 |
| P | | < .01† | |
| *Novikoff hepatoma* | | | |
| 11 | 155 | 0.798 | 0.120 |
| 12 | 160 | .852 | .120 |
| 13 | 231 | .827 | .115 |
| Mean and S.E. | | 0.826 ± 0.013 | 0.118 ± 0.002 |
| P | | < .01† | |
| *Fibroadenoma (benign)* | | | |
| 14 | | 0.448 | |
| 15 | | .537 | |
| Mean | | .492 | |

Damadian published his results ahead of us because he had a simpler objective, namely, he wanted to show that the T1 of cellular water is longer in some tumors in comparison to normal tissues in general. In his 1971 paper, he did clearly acknowledge that his work was inspired by the findings of our team [31]. In our case, our study was aimed at testing whether NMR can be used as a reliable method to detect cancer. We need to conduct a more complete/detailed study to demonstrate that. For example,

(1) Unlike Damadian who measured the T1 in three different tumors and compared them with various normal tissues samples, we measured the NMR properties of cellular water in the same tissue model (breast tumor) at three different morphological states (normal tissue, pre-neoplastic nodule, and cancer). So, we can clearly correlate the changes of NMR parameters with different stages of cancer development [27, 28].

(2) We measured not only the spin-lattice relaxation time T1, but we also measured the spin-spin relaxation time T2 and the spin diffusion coefficient (D) of water protons. We found all three NMR parameters undergoing a progressive change during cancer development [27, 28].

(3) We repeated our measurements in sufficiently large number of samples to allow a clear statistical test on the significance of the experimental findings.

(4) A major discovery of our study was that the NMR properties of water proton start to undergo distinct changes during the pre-neoplastic stage. **This suggests NMR can be used as an early diagnostic tool**, so that medical treatment can be applied before cancer formation [27, 28].

After our finding was reported in the APS March Meeting, several groups became interested in using NMR to study biological tissues, including tumors [32–34]. One of these groups was the team led by Weisman, who studied the T1 changes of a rat tail during tumor development [32]. Their result in general confirmed what we found earlier in the NMR study of mammary tumor [27, 28].

The reports by Damadian and our group also stimulated D. Hollis to use NMR to study tumors [33]. Hollis was skeptical about the claim of Damadian that one can use the value of T1 alone to differentiate tumors from non-tumor cells. So, Hollis decided to repeat Damadian's study; his team used the same pulsed NMR spectrometer owned by **NMR Specialties** that Damadian had used before [33]. Hollis concluded that, although their results confirm the earlier studies, he suspected that abnormal states other than cancer might produce elevated T1 values. Particularly, he noticed that an adenocarcinoma of the lung was not significantly different in its T1 value from adjacent uninvolved lung [33, 35].

The study by Hollis in fact testified to the importance of using multiple NMR parameters to develop NMR as a diagnostic tool. That was why we investigated the changes of T1, T2 and the spin diffusion coefficient D during cancer development in our earlier study [27, 28]. In our case, there was clear statistical significance to show that NMR measurement can be used to distinguish the three different morphological states in mammary tumor (i.e., normal tissue, pre-neoplastic nodule, and cancer) [27, 28].

Following our reports of 1972, our lab continued to engage actively in subsequent studies of developing NMR for detecting cancer [36]. Particularly, we further showed that the NMR method can be applied to differentiate between normal, pre-cancer, and cancer cells in **human breast tissue**. Our results indicated that NMR relaxation times could distinguish between the breast neoplasms and other diseased or normal tissues with $P$ values <0.001 [36]. For a given sample, the probability of classifying it nonneoplastic or carcinoma could be accomplished with 85% confidence. Furthermore, the relaxation time T2 was found to be more discriminating than T1 in detecting cancer in human breast tissues [36].

## The use of magnetic field gradient to generate 2-D NMR images

In the early 1970s, imaging techniques using X-ray (CT) had become widely known [37]. Some scientists started to think about applying a magnetic field gradient to generate 2-D image from NMR measurement [38–40]. The principle is very simple. Since the precession frequency of a nuclear spin is proportional to the local magnetic field, when one applies a magnetic field gradient to the sample, the nuclei located at different positions of the sample will process at different frequencies; their NMR signal can be separated by using Fourier de-composition.

One year after we reported our NMR study of mammary tumor at the APS March Meeting in 1972, Paul Lauterbur, a chemistry professor and an executive of the **NMR Specialty Corporation**, published a short paper in *Nature* to propose using a fixed magnetic field gradient to generate a 2-D NMR image. He called his method "*zeugmatography*" [38]. This study was not directly related to biology or medicine. The sample used by Lauterbur was two tubes of $H_2O$. By applying a magnetic field gradient to scan the sample along one axis, and then rotating the sample for scanning along other axes, he generated a 2-D image of the distribution of water proton using CW NMR. (See **Fig. 2**).

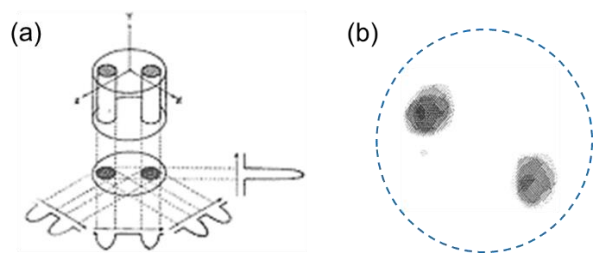

**Fig.2.** (a) The zeugmatography method proposed by Lauterbur in 1973. He used a fixed magnetic field gradient to measure the proton density of the sample projected on one plane. Then, by rotating the orientation of the field gradient, other projections can be obtained. (b) The 2-D image produced by Lauterbur, for a sample consisting of two tubes of pure water. (Reproduced from Ref [38]).

At about the same time, a British physicist (Peter Mansfield) was also interested in applying a magnetic field gradient to generate a 2-D scanning image for solid state physical samples [39]. Within a few years, several groups had claimed to successfully generate magnetic images of animal or human body [41–44]. (See **Fig. 3**). One of the active players in this period was Damadian. He invented a magnetic imaging method called "FONAR", which used a specially designed magnetic field gradient [42, 45]. Damadian filed a patent application in 1972 proposing to build a whole-body NMR machine for cancer detection [46]. He also published the first human thoracic image using NMR in 1977 [45].

These works stimulated many other laboratories to rush into the development of magnetic imaging. Most of the images were based on NMR measurements of the spin density, i.e., concentration of water protons. These images were very crude; they were published mainly for the purpose of competing for priority record.

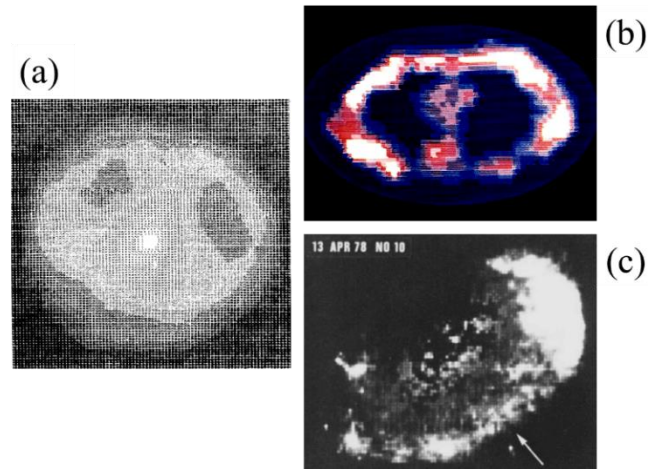

**Fig. 3.** Early NMR images in the 1970s. (a) A proton NMR image of the thoracic cavity of a mouse, as shown by Lauterbur in ref [41]. (b) A FONAR image of human chest, as shown by Damadian in ref [42]. (c) A line-scan NMR image of the human abdomen, as shown by Mansfield in ref [43].

The focus of research at that time was to device more sophisticated magnetic field gradients to improve the 2-D scanning image. Different designs of magnetic field gradient (G) had been attempted, including, static G vs time-variating G, linear G vs non-linear G, and single G along one axis vs multiple Gs along different axes. Some of these designs were found to be useful in generating NMR images of improved quality [39–44].

## MRI based on relaxation time measurements using spin-echo NMR

Up to the beginning of 1980s, most NMR images on biological bodies were generated from measurements of nuclear spin density, which gave the distribution of water protons in the sample. The NMR images at that time were far less clear than the X-ray images using CT and could not give sufficient anatomical details. Thus, the NMR imaging technique was only an experimental tool with limited uses.

The situation changed starting from the early 1980s, when the new generation of NMR imaging device began to use spin-echo NMR to measure the relaxation times of tissue water at different locations of the biological sample [47, 48]. As discovered in the early 1970s, the NMR relaxation times of the water protons are sensitive to the physiological-pathological states of the tissue [19, 22, 23]. So, **the NMR images generated from the relaxation time measurement can show whether the tissue is normal or in a pathological state**, particularly, whether the tissue has become cancer or not. This ability has great diagnostic value and can make the NMR imaging more useful than the CT scan.

By this time, the medical radiologist community decided to change the name "*NMR imaging*" to "*Magnetic Resonance Imaging*" (MRI). This is to avoid the association with "*nuclear radiation*", which could generate a bad feeling in some patients.

Starting from the early 1980s, MRI had become a very active research area in medical physics. Some of the major hospitals began to equip them with MRI machines. Good quality images were routinely produced (see **Fig. 4**). MRI had become an important tool for the diagnostic radiologists [48].

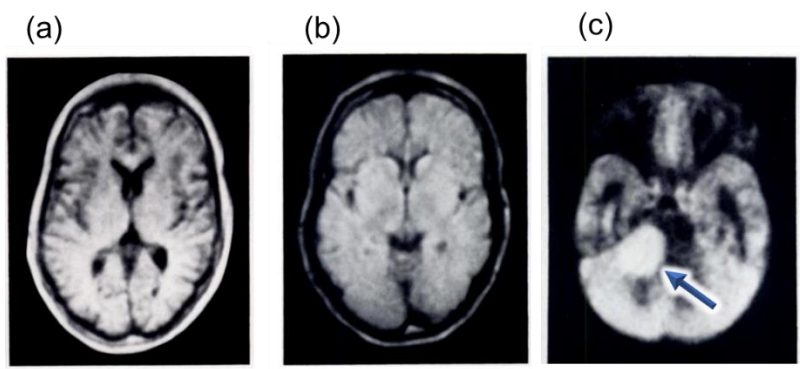

**Fig. 4.** (a-b) MRI image of normal human brain at low ventricular level: (a) MRI image based on T1 measurements; (b) MRI image based on T2 measurements. (c) MRI image showing a human brain tumor (acoustic neuroma), see arrow. (Reproduced from Ref. [48]).

The development of MRI after 1980s was further benefitted from several additional technical advancements. First, the availability of affordable powerful computers for image processing. Both the hardware and software of computers were developed almost at an exponential rate during this time. Second, with the rapid advancement of the computer power, it becomes practical to develop the "spin warp" technique, which can greatly speed up the MRI measurement process [49, 50]. Third, the supply of superconducting magnets with large bore; it can generate a homogeneous magnetic field with sufficient volume to fit in a patient. Finally, the development of paramagnetic agents (such as Gd-based contrast agents) which can greatly enhance the contrast of the MRI images [51].

## Discussions

Nowadays, MRI is routinely used to detect tumors in patients. It can give remarkably clear anatomical details [52]. MRI has become a widely used method for radiologists to diagnosis whether a patient may have cancer or not [52].

Lauterbur was very active in promoting magnetic imaging during the 1970s [5, 11, 53, 54]. This gave people the impression that he was the inventor of using NMR for cancer detection. Particularly, in his 1973 *Nature* paper, he omitted to mention Damadian's work or our work. (Note: He did cite a paper by Weisman *et.al.*, which cited our 1972 APS report [27, 32].) This resulted in a common misunderstanding that Lauterbur was the principal inventor of MRI for medical uses today [8].

When Damadian found out that Lauterbur had not cited his work, he expressed doubts over Lauterbur's intentions [53]. According to a review written by A. Prasad, Damadian claimed "that Lauterbur was trying to steal his ideas, which were known to the employees of **NMR Specialties**, where Lauterbur had worked when he came up with the *zeugmatography* method" [5]. From then on, the two became bitter enemies and engaged in a hot race for priority.

Damadian established a commercial company (the FONAR Corporation) in late 1970s. In 1977, Damadian's group used their FONAR method to produce what is considered the first NMR image of the human chest. As pointed out by Prasad, "both Lauterbur and Damadian moved between different 'social worlds' in order to develop MRI and to strengthen their priority claims for its invention…

Lauterbur was as interested in gaining priority as Damadian was, and he too moved across different social worlds to strengthen his claim" [5].

However, according to Prasad, Damadian flouted the conventions of the scientific community "when he asked politicians to intervene in the funding process and when he announced his group's results via a news conference even before he had presented them in a scientific forum. Damadian's actions earned him a 'bad boy' image within the scientific community. Nevertheless, Damadian was able to arouse enormous interest in the media for NMR imaging" [5]. And at the same time, Lauterbur energetically took his idea to scientific meetings in the US and abroad. "He became a one-man traveling evangelical show, teaching the new religion of NMR imaging" [53].

Subsequently, Lauterbur was awarded the Nobel Prize of Physiology/Medicine in 2003. This further reinforced the belief that Lauterbur's *zeugmatography* was the precursor of the diagnostic MRI used today [14]. However, this understanding is not completely true if one examines the published record carefully. First, the *zeugmatography* was based on CW NMR, not spin-echo NMR. Second, the *zeugmatography* measured the distribution of spin density in the sample, not the position-dependence of water proton relaxation times. Finally, the 2-D scanning of spin density obtained in the original *zeugmatography* was not aimed for distinguishing cancer from normal tissue. The **key finding for developing MRI as a diagnostic tool for cancer** was the discovery that **the NMR relaxation times (T1, T2) and the spin diffusion coefficient D are sensitive to the physiological-pathological state of the biological tissues** [22, 23, 26].

Therefore, Damadian thought that he should be more deserving of the credit of developing MRI [9, 13, 45, 53]. Today, the MRI community appears to be highly polarized between the Lauterbur side and the Damadian side [7–12, 53, 54]. Most late comers in this field just could not spend the time to examine the publication record carefully. My understanding is that Damadian and Lauterbur had made different contributions in this field. As we showed earlier in this paper, there were three different major contributions for the development of MRI, (a) the use of spin-echo NMR for relaxation time measurement in biological tissues; (b) the discovery that T1, T2, and the spin-diffusion coefficient D are sensitive to the physio-pathological state; and (c) the use of magnetic field gradient for 2-D scanning. As far as I know, no one single scientist contributed all three and with the earliest priority. The development of MRI for cancer detection is a triumph of the study of medical physics, which was contributed by many scientists. The argument between Lauterbur and Damadian was somewhat misleading. Their intense fighting in public actually had overshadowed the contribution of other scientists in this field.

From the published research records, one can see that three major groups (including Damadian, the Baylor/Rice team, and Lauterbur) had made important contributions in the development of MRI for cancer detection. **Table 4** is a short summary of major contributions from these three groups.

**Table 4: A Short Summary of the Early History of Using NMR for Cancer Detection**

| Research Team | Contribution/findings | Remark |
|---|---|---|
| **C.F. Hazlewood** and **D.C. Chang** (CW-NMR 1969-1970; spin-echo NMR 1970- ) | Summary of works from 1969 to 1972: [15, 22, 25–30]<br>(1) The NMR relaxation times T1, T2 and spin diffusion coefficient D were found to be very different from those of bulk water, implying some sort of ordering in the cellular water.<br>(2) The degree of ordering of the tissue water seems to depend on the differentiation state of the tissue; thus, T1 and T2 were longer in the less differentiated (i.e., immature) tissues in comparing to mature tissue.<br>(3) As expected, tumor development is associated with de-differentiation; they found the relaxation times T1 and T2 (as well as D) became significantly lengthened during tumor development in a specific tissue (mammary gland).<br>(4) Particularly, the NMR measurements were able to clearly distinguish the pre-tumor tissue from the normal tissue or tumor, suggesting that NMR can be used as an early diagnostic tool, so that medical treatment can be applied before cancer formation. | • CFH used CW-NMR to study the state of tissue water before 1970 [22, 23]<br>• In 1970, DCC joined in by converting his home-built spin-echo NMR apparatus for studying the physical state of water protons in biological tissues [15, 24–28] |
| **R. Damadian** (spin-echo NMR 1971- ) | As reported in his 1971 *Science* paper: [31]<br>(1) He measured T1 of water proton in six normal tissues in the rat and in two malignant solid tumors.<br>(2) Relaxation times for the two malignant tumors were distinctly different from the normal tissues.<br>(3) The T1 for benign fibroadenomas were distinct from those malignant tissues and were the same as those of muscle.<br>Implications: NMR can be used to differentiate malignant tumors from normal tissues; but it was unclear whether benign tumors can be differentiated. | • RD published his first NMR paper in 1971.<br>• RD acknowledged that this study was inspired by CFH's earlier findings and personal communication. [31]<br>• RD filed a patent application in 1972 proposing to build a NMR machine for cancer detection [46] |
| **P. Lauterbur** (CW-NMR; 1973- ) | As reported in his 1973 *Nature* paper: [38]<br>(1) NMR zeugmatography was performed with 60 MHz radiation and a static magnetic field gradient generator. The sample was two tubes of $H_2O$.<br>(2) The application of a magnetic field gradient allows the sample to be scanned along one axis; by rotating the sample, scanning along other axes can also be achieved.<br>(3) Thus, one can generate a 2-D image of the distribution of a specific type of nuclei (water proton) using NMR | • His method mainly measured the distribution of nuclei rather than the physical state of the nuclei.<br>• No biological tissue was involved in this study.<br>• He cited a paper by Weisman et al, which cited the 1972 *APS abstract* by Chang et al. [27, 32, 38] |

From my experience in the study of NMR for cancer detection, I strongly feel that physics is a wonderful discipline that has great potential to benefit mankind. Not only physics provides the scientific basis for the industrial revolution, which greatly improved our living standard, physics can also contribute to the upkeeping of human health. MRI is just one good example. We are also reminded that important technologies are usually developed based on multi-disciplinary research. For example, the key discovery involved in MRI was contributed by scientists from multiple disciplines, including physics, chemistry, physiology, and pathology.

**Acknowledgement**:

I thank Drs. Stephen Russek, Richard Spencer, and Jenifer Pursley for useful feedback. I also thank Ms. Lan Fu for her assistance. This work is partially supported by grants from RGC of Hong Kong (RMGS20SC01) and HKUST (DCC18SC01).